# Light Field Distributions in One-dimensional Photonic Crystal Fibers

**Changbiao Wang**
*ShangGang Group, 70 Huntington Road, Apt. 11, New Haven, CT 06512, USA*
*changbiao_wang@yahoo.com*

It is shown in this paper that the light field distribution in a band gap within periodic structures for one-dimensional photonic crystal fibers is described by a decaying factor multiplied by a periodical function that has the same period length as the one of the medium or has double the period length of the medium, depending on the sign of the trace of eigen value matrix. This fundamental property is applicable to any 1D planar periodic structures, no matter how many layers a unit cell has, what the contrast of refractive indices is, and whether the dielectric parameters in individual layers are homogeneous or inhomogeneous; it plays a significant role in understanding of numerical results in a number of previously published research works. It is also shown that, similar to the refractive index guidance in conventional optical fibers, the photonic band gap guidance is also a form of total internal reflection.

## 1. INTRODUCTION

Photonic crystals are artificial materials with periodic variations in refractive index that are designed to affect the propagation of light [1,2,3,4]. An important feature of the photonic crystals is that there are allowed and forbidden ranges of frequencies at which light propagates in the direction of index periodicity. Due to the forbidden frequency range, known as photonic band gap (PBG) [5,6], the photonic crystals can be used to form PBG mirrors for waveguiding, resulting in a new kind of optical fibers, where light with frequencies falling in the PBG is confined in a hollow core surrounded by layered dielectrics [7,8] or by a array of microscopic air holes [9]. Unlike the conventional optical fibers which guide light by total internal reflection through a solid core which has a higher refractive index (index guidance), the photonic crystal fibers guide the light by PBG effect through a hollow core which has a lower index (PBG guidance) [10,11], thus leading to lower losses and allowance of fibers to take sharp bends [12,13].

One-dimensional (1D) PBG fibers, consisting of a guiding region and two planar periodically stratified uniform media, were first analyzed in the pioneering works [14,15], where the authors called such structures Bragg reflection waveguides. To improve the band gap, various profiles of refractive index have been suggested and analyzed [16,17]. In the 1D PBG fibers, the field distributions of a guided mode in the periodic structures are characterized by a Floquet solution (Bloch wave); thus Floquet's theorem (Bloch analysis) is a fundamental tool for analysis of such structures. Although computations are highly developed, analytic theoretic results never become less important because they can be used to help understand the results from computations [18]. Based on a complex translation-matrix method, Yeh, Yariv, and Hong first applied Floquet's theorem to analysis of the 1D PBG fibers, showing a strong localization of field distributions in the guiding region. In their theory, the eigen value matrix is complex [15], and the property of the reflection coefficient in periodic structures to be obtained is not easy to directly identify. In this paper, a real translation-matrix method is employed to derive a general concise expression of reflection coefficient. A straightforward analysis of this expression indicates that the PBG guidance is also a form of total internal reflection. It is also shown by a further analysis of Floquet's theorem that the light field distribution in a band gap within the periodic structures is characterized by a decaying factor multiplied by a periodical function that has the same period length as the one of the medium if the band index is even, or has double the period length of the medium if the band index is odd. This simple and fundamental property is confirmed by numerical examples based on the formulism developed in this paper, and also plays a significant role in understanding of numerical results in a number of previously published research works.

The paper is organized as follows. In Sec. 2, general equations for 1D PBG fibers are given and a construction of Floquet solutions (Bloch waves) is described. In Sec. 3, by using a real translation-matrix method, a general expression of reflection coefficient is derived. In Sec. 4, the properties of Floquet solutions (Bloch waves) are analyzed, and it is shown that the PBG guidance is another form of total internal reflection. Finally, a summary is given in Sec. 5.

## 2. GENERAL EQUATIONS FOR 1D PHOTONIC CRYSTAL FIBERS

In this section, general equations for loss-free 1D photonic crystal fibers are given and a construction of Floquet solutions (Bloch waves) is described.

We assume that the permittivity $\varepsilon$ and the permeability $\mu$ in a 1D dielectric medium are only functions of $x$ in space, namely $\nabla \varepsilon \equiv \hat{\mathbf{x}} \partial \varepsilon / \partial x$, and $\nabla \mu \equiv \hat{\mathbf{x}} \partial \mu / \partial x$, where $\hat{\mathbf{x}}$ is the unit vector in the $x$-direction. The electric and magnetic fields $(\mathbf{E}, \mathbf{H})$ have a time-space wave factor $\exp[i(\omega t - k_z z)]$ and have no variations in the $y$-direction [ $\partial(\mathbf{E}, \mathbf{H})/\partial y \equiv 0$ ], where $\omega$ is the angular frequency and $k_z$ is the axial wave number. For TM waves ($H_x = 0$), corresponding to LSM modes in rectangular dielectric-loaded waveguides [19], the electric field $E_x$ satisfies

$$\frac{\partial}{\partial x}\left(\varepsilon \frac{\partial E_x}{\partial x}\right) + \varepsilon\left[(\omega^2 \mu \varepsilon - k_z^2) + \frac{d}{dx}\left(\frac{1}{\varepsilon} \frac{d\varepsilon}{dx}\right)\right] E_x = 0, \tag{1}$$

and the rest components are given by

$$E_z = \frac{1}{ik_z}\left(\frac{\partial E_x}{\partial x} + \frac{1}{\varepsilon}\frac{d\varepsilon}{dx} E_x\right), \qquad H_y = \frac{\omega \varepsilon}{k_z} E_x, \tag{2}$$

and $E_y = 0$, and $H_x = H_z = 0$. If $k_z = 0$ (perpendicularly incident wave), $E_x$ must be equal to zero. To get a non-trivial solution, use $(k_z c/\omega)E_x$ to replace $E_x$, with $c$ the vacuum light speed, and then take a limit, namely $(E_x, E_z, H_y) = \lim(E_x, E_z, H_y)$ as $k_z \to 0$.

For TE waves ($E_x = 0$), corresponding to LSE modes in rectangular dielectric-loaded waveguides, the magnetic field $H_x$ satisfies

$$\frac{\partial}{\partial x}\left(\mu \frac{\partial H_x}{\partial x}\right) + \mu\left[(\omega^2 \mu \varepsilon - k_z^2) + \frac{d}{dx}\left(\frac{1}{\mu} \frac{d\mu}{dx}\right)\right] H_x = 0, \tag{3}$$

and the rest components are given by

$$H_z = \frac{1}{ik_z}\left(\frac{\partial H_x}{\partial x} + \frac{1}{\mu}\frac{d\mu}{dx} H_x\right), \qquad E_y = -\frac{\omega \mu}{k_z} H_x, \tag{4}$$

and $E_x = E_z = 0$, and $H_y = 0$. If $k_z = 0$, a similar treatment is needed. In fact, all field components for TE waves can be directly obtained from TM waves by substitutions of $E \to H$, $H \to E$, $\varepsilon \to -\mu$, and $\mu \to -\varepsilon$, because the system of source-free Maxwell's equations are the same under these substitutions.

TM and TE waves have no coupling although $\varepsilon$ and $\mu$ are both functions of $x$. Since they only have one generating component ($E_x$ for TM and $H_x$ for TE), there are only two boundary conditions to match on each dielectric interface [19].

For loss-free 1D photonic crystal fibers, $\varepsilon(x)$ and $\mu(x)$ are real and have the same periodicity. It is seen from Eqs. (1) and (3) that $E_x$ and $H_x$ satisfy the same kind of equation in the form of

$$\frac{d}{dx}\left[p(x)\frac{dy}{dx}\right] + q(x; \omega/c, k_z) y = 0, \tag{5}$$

where $p(x)$ and $q(x; \omega/c, k_z)$ are periodic real functions, namely $p(x + \Lambda) = p(x)$ and $q(x + \Lambda) = q(x)$, with $\Lambda$ the minimum period of medium.

If a function, $y(x)$, is a solution to Eq. (5) and its function values at any two locations separated by one period are different by a so-called real or complex characteristic constant (eigen value), $\sigma$, that is,

$$y(x + \Lambda) = \sigma y(x), \tag{6}$$

then $y(x)$ is said to be Floquet solution (Bloch wave) [20]. Floquet solutions are the solutions that satisfy both Eq. (5) and Eq. (6). The linear combination of two linearly independent Floquet solutions constitutes a general solution; of course, the general solution does not have to be composed of Floquet solutions. Unless $\alpha\beta = 0$, any linear combination of two linearly independently Floquet solutions, $\alpha y_1(x) + \beta y_2(x)$, is not a Floquet solution any more; in other words, a Floquet solution is a particular solution.

Suppose that $f(x)$ and $g(x)$ are two linearly independent solutions (basis solutions) [21]. A Floquet solution can be written as $y(x) = C_1 f(x) + C_2 g(x)$, and seeking Floquet solutions (Bloch waves) is actually to determine the ratio of $C_1$ to $C_2$ so that the imposed physical condition $y(x + \Lambda) = \sigma y(x)$ is satisfied.

The basis solutions $f(x)$ and $g(x)$ can be real or complex. All complex forms of basis solutions, of which the real and imaginary parts must satisfy Eq. (5), can be taken to be the result from real basis solutions by linear transforms, and the eigen value $\sigma$ is an invariant of the linear transforms. This property can be used for simplifying calculations.

## 3. REFLECTION COEFFICIENT IN 1D PBG STRUCTURES

In this section, with the help of a real translation-matrix method, a general concise expression of reflection coefficient on the dielectric interfaces is derived for a periodically stratified uniform medium of which the unit cell has any number of layers.

It is seen from Eqs. (1) and (3) that TM and TE waves for a stratified uniform medium satisfy the same much-simplified wave equation due to $d(\varepsilon,\mu)/dx = 0$, but they have different boundary conditions to match on dielectric interfaces. Nevertheless, the derivations of the reflection coefficients for the two waves are similar and only the one for TM waves is given below.

Suppose that the electric field component perpendicular to the dielectric interface can be written as $E_x(x,z) = E(x)e^{i(\omega t - k_z z)}$, where $E(x)$ for the $n^{th}$ layer is given by

$$E(x) = a_n e^{-ik_x^{(n)}x} + b_n e^{+ik_x^{(n)}x}, \qquad (x_{n-1} < x < x_n). \qquad (7)$$

In the above, $a_n$ and $b_n$ are constants, $x_n - x_{n-1} = \Lambda_n$ is the layer's thickness, and $(k_x^{(n)})^2 = \omega^2 \varepsilon^{(n)} \mu^{(n)} - k_z^2$. It is assumed that there are $m$ layers in one period (one unit cell): $x_{n+m} - x_n = \Lambda$, $\varepsilon^{(n+m)} = \varepsilon^{(n)}$, $\mu^{(n+m)} = \mu^{(n)}$, and $k_x^{(n+m)} = k_x^{(n)}$.

To construct the Floquet solution from the general solution given by Eq. (7), the basis solutions can be taken to be $f(x) = \exp(-ik_x^{(n)}x)$ and $g(x) = \exp(+ik_x^{(n)}x)$, with $y(x) = E(x)$, $C_1 = a_n$, and $C_2 = b_n$ following the description in Sec. 2. Once the ratio $b_n/a_n$ is determined, the Floquet solution and the reflection coefficient are obtained.

Imposing Floquet condition Eq. (6) on Eq. (7), we have

$$\begin{pmatrix} a_{n+m} \\ b_{n+m} \end{pmatrix} = \sigma (U^T)^{-1} \begin{pmatrix} a_n \\ b_n \end{pmatrix}. \qquad (8)$$

where $U = \mathrm{diag}(e^{-ik_x^{(n)}\Lambda},\ e^{+ik_x^{(n)}\Lambda})$.

The above Eq. (8) is a sufficient and necessary condition of the existence of Floquet solution for a periodically stratified uniform medium. As we know, boundary conditions also result in another equation. To get a real layer's translation matrix, let us define a vector, which is continuous at all dielectric boundaries, given by [19]

$$\begin{pmatrix} \varepsilon E(x) \\ E'(x) \end{pmatrix} = M_n(x) \begin{pmatrix} a_n \\ b_n \end{pmatrix}, \qquad (9)$$

where $E'(x) \equiv dE(x)/dx$ is proportional to $E_z$ as shown in Eq. (2), and

$$M_n(x) = \begin{pmatrix} \varepsilon^{(n)} e^{-ik_x^{(n)}x} & \varepsilon^{(n)} e^{+ik_x^{(n)}x} \\ -ik_x^{(n)} e^{-ik_x^{(n)}x} & ik_x^{(n)} e^{+ik_x^{(n)}x} \end{pmatrix}. \qquad (10)$$

Applying the continuity of $(\varepsilon E, E')$ at $x = x_n$ to Eq. (9), we obtain a recurrence expression, given by

$$M_n(x_n) \begin{pmatrix} a_n \\ b_n \end{pmatrix} = M_{n+1}(x_n) \begin{pmatrix} a_{n+1} \\ b_{n+1} \end{pmatrix}. \qquad (11)$$

From above, we obtain

$$M_{n+m}(x_{n+m}) \begin{pmatrix} a_{n+m} \\ b_{n+m} \end{pmatrix} = T M_n(x_n) \begin{pmatrix} a_n \\ b_n \end{pmatrix}, \qquad (12)$$

or

$$\begin{pmatrix} \varepsilon E(x_{n+m}) \\ E'(x_{n+m}) \end{pmatrix} = T \begin{pmatrix} \varepsilon E(x_n) \\ E'(x_n) \end{pmatrix}, \qquad (13)$$

where $T = T_{n+m} \cdots T_{n+2} T_{n+1}$, with $T_n = M_n(x_n) M_n^{-1}(x_{n-1})$. $T$ is the one-period translation matrix of $(\varepsilon E, E')^T$ on dielectric boundaries while $T_n$ is its one-layer translation matrix, which is real, given by

$$T_n = \begin{pmatrix} \cos\phi_n & (\varepsilon^{(n)}/k_x^{(n)})\sin\phi_n \\ -(k_x^{(n)}/\varepsilon^{(n)})\sin\phi_n & \cos\phi_n \end{pmatrix}, \tag{14}$$

with $\phi_n = k_x^{(n)}\Lambda_n$. This $T_n$-matrix also can be obtained from an alternative treatment where the two field components, parallel to the dielectric interface, are taken to be a set of basis solutions, as shown by Lekner [22].

Inserting Eq. (8) into Eq. (12), we obtain the eigen matrix equation and eigen value equation, given by

$$(T - \sigma I) M_n(x_n) \begin{pmatrix} a_n \\ b_n \end{pmatrix} = 0, \tag{15}$$

$$\sigma^2 - \sigma(t_{11} + t_{22}) + 1 = 0 \tag{16}$$

where $I$ is the unit matrix, and $|M_n(x_n)| \neq 0$ and $M_{n+m}(x_{n+m})(U^T)^{-1} = M_n(x_n)$ are employed. It is seen from Eq. (16) that the eigen value $\sigma$ only depends on the trace $t_{11} + t_{22}$, a sum of diagonal elements of the eigen value matrix $T$, and the trace is real and independent of the choice of basis solutions.

Inserting the eigen value $\sigma$ obtained from Eq. (16) into Eq. (15), we get $b_n/a_n = \Gamma(x_n)e^{-i2k_x^{(n)}x_n}$, where $\Gamma(x_n)$ is the general expression of reflection coefficient on dielectric interfaces, given by

$$\Gamma(x_n) = -\frac{\varepsilon^{(n)}(t_{11}-\sigma) - i(k_x^{(n)}t_{12})}{\varepsilon^{(n)}(t_{11}-\sigma) + i(k_x^{(n)}t_{12})}, \tag{17}$$

or

$$\Gamma(x_n) = -\frac{\varepsilon^{(n)}t_{21} - ik_x^{(n)}(t_{22}-\sigma)}{\varepsilon^{(n)}t_{21} + ik_x^{(n)}(t_{22}-\sigma)}. \tag{18}$$

In the above, $t_{11}$, $t_{12}$, $t_{21}$, and $t_{22}$ are all the elements of the eigen value matrix $T$, and they are real. Clearly, the numerator and denominator of the right-hand sides of the above equations are conjugate complex if $\sigma$ and $k_x^{(n)}$ are real, leading to $|\Gamma(x_n)| = 1$ or $|b_n/a_n| = 1$ except for those that lead to the indeterminate form 0/0 taking place from *both* Eqs. (17) and (18). Considering that $T$ is a unimodular matrix ($|T|=1$), we obtain an important result, $a_n a_n^* - b_n b_n^* = 0$ for all real $\sigma \neq \pm 1$ and $k_x^{(n)} \neq 0$.

It is seen from Eq. (14) that the one-period translation (eigen value) matrix $T$ has a noticeable advantage: $T$ is the product of individual layer's $T_n$-matrices that only depend on individual layer's dielectric parameters and thickness, while in the traditional derivation the translation matrix is made up of individual matrices that contain two adjacent layer's dielectric parameters [15]. Another advantage is that $T$ is a real matrix even if $k_x$ is imaginary [19], while in the classic stratified-medium treatment by Abelès, the anti-diagonal elements of such a translation matrix are imaginary for a loss-free medium [23]. Probably, using a real matrix to characterize transmission of a loss-free medium is better matching with the custom of using real permittivity and permeability to describe dielectric property of such a medium.

A continuously varying medium can be taken to be the limit of a corresponding stratified uniform medium as long as all layer's dielectric parameters follow given profiles and the layer's number is large enough. Since $T = T_{n+m} \cdots T_{n+2} T_{n+1}$ is real, independently of the number of layers in a unit cell to be set, it can be inferred that the eigen value matrix is also real for a general periodically stratified medium where the dielectric parameters may be continuously or partially continuously varying in individual layers [16,17].

It should be pointed out that the eigen value matrix $T = T_{n+m} \cdots T_{n+2} T_{n+1}$ depends on the layer's sequence number $n$, and there may be $m$ different eigen value matrices for a periodic structure with a unit cell having $m$ layers. But all the eigen value matrices have the same eigen values, because these eigen value matrices are just the products of different cyclic permutations of all $m$ layer's translation matrices [24].

## 4. PROPERTIES OF FLOQUET SOLUTIONS (BLOCH WAVES)

It has been well realized that the Floquet solution (Bloch wave) in a band gap can be described by a decaying factor multiplied by a periodic function [14,15,25]; however, it has never been analytically shown what period the function has and what the period length depends on. In this section, this issue is addressed by a further analysis of Floquet's theorem.

By setting $\sigma = e^{-iK\Lambda}$ to convert $\sigma$ into the characteristic wave number $K$, from Eq. (6) the Floquet solution (Bloch wave) can be written as [26]

$$y(x) = e^{-iKx}\psi_K(x), \tag{19}$$

where $\psi_K(x) = \psi_K(x+\Lambda)$ is a periodic function. Inserting $\sigma = e^{-iK\Lambda}$ into Eq. (16), we obtain the well-known *K*-equation

$$\cos(K\Lambda) = \frac{1}{2}(t_{11} + t_{22}). \qquad (20)$$

For given $k_0 = \omega/c$ and $k_z$ in a band gap, the Floquet solution obtained is made up of a decaying factor multiplied by a periodic function, termed *decaying periodic solution* for the convenience. Suppose that the complex characteristic wave number can be written to be $K = K_r + iK_i$, where $K_r$ and $K_i$ are all real. Since the trace $(t_{11} + t_{22})$ is real, from Eq. (20), $K_r = l\pi/\Lambda$ must hold for $K_i \neq 0$, where $l$ is an integer, termed band index [15]; thus leading to a real $\sigma = (-1)^l e^{K_i\Lambda} \neq \pm 1$. Therefore, the *K*-equation can be re-written as

$$K_r = l\pi/\Lambda, \quad (-1)^l \cosh(K_i\Lambda) = \frac{1}{2}(t_{11} + t_{22}), \qquad (K_i \neq 0). \qquad (21)$$

No mater whether *l* is odd or even, $|(t_{11} + t_{22})/2| > 1$ must hold and the two roots for $K_i$ can be written as

$$K_i = \pm \frac{1}{\Lambda} \ln\left( \left|\frac{1}{2}(t_{11} + t_{22})\right| + \sqrt{\frac{1}{4}(t_{11} + t_{22})^2 - 1} \right). \qquad (22)$$

From Eqs. (19), (21), and (22), the two linearly independent Floquet solutions can be written as

$$y(x) = e^{\pm|K_i|x} e^{-il\pi x/\Lambda} \psi_K(x). \qquad (23)$$

It follows from Eqs. (21) and (23) that there are two kinds of Floquet solutions (Bloch waves), depending on the sign of the trace $t_{11} + t_{22}$, as shown below:

1. *Decaying half-period Floquet solution*. If $(t_{11} + t_{22}) < -2$, then from Eq. (21) the band index *l* must be an odd integer. The factor $e^{-il\pi x/\Lambda}\psi_K(x)$ in Eq. (23) is a function that has a minimum period of $2\Lambda$, that is, double the medium period; we call it a half-period function because it runs one half period over a full period of the medium.

2. *Decaying full-period Floquet solution*. If $(t_{11} + t_{22}) > +2$, then the band index *l* must be an even integer. In such a case, the factor $e^{-il\pi x/\Lambda}\psi_K(x)$ in Eq. (23) is a function that has a minimum period of $\Lambda$; we call it a full-period function because it has the same period as the one of the medium.

It is seen from Eq. (21) that a collection of all $k_0$ and $k_z$ in the regime $|\cos(K\Lambda)| = \cosh(K_i\Lambda) = |(t_{11} + t_{22})/2| > 1$ constitutes a band gap, while $|t_{11} + t_{22}| - 2 = 0$ forms its boundary. Out of the band gap, $|\cos(K_r\Lambda)| \leq 1$ ($K_i = 0$) is an allowed band, Floquet solutions correspond to propagating modes, and they may be periodical or not periodical. For example, when $K_i = 0$, $K_r = \pi/\Lambda$ and $2\pi/\Lambda$ correspond to stable half- and full-period Floquet solutions respectively, while $K_r = \sqrt{2}\pi/\Lambda$ corresponds a Floquet solution that is not periodical because $\exp(-i\sqrt{2}\pi x/\Lambda)$ and $\psi_K(x)$ do not have a common period.

Physically, Floquet solutions (Bloch waves) in the band gap exist only in a matched half-infinite periodic structure because a Floquet solution (Bloch wave) is a particular solution as mentioned in Sec. 2 and it only has one solution coefficient to be determined. Such a half-infinite structure can be taken to be an approximation of a practical structure that is long enough so that the electromagnetic fields have almost vanished at the structure end. Several schemes for supporting Floquet solutions (Bloch waves) used for 1D photonic crystal fibers were proposed and analyzed [14,15], and in recent years, photonic crystal fibers have been suggested for particle acceleration [27,28,29], where a vacuum-light-speed mode may be supported. According to the conclusion obtained above, there are two kinds of Floquet solutions (Bloch waves) in the band gap: decaying half-period solution and decaying full-period solution. Such numerical examples can be found in the published works mentioned above.

(1). *Numerical examples for decaying half-period Floquet solutions* ($2\Lambda$). Figures 2 and 3 of Ref. 14, figures 19, 20, and 22 of Ref. 15, figure 4 of Ref. 27, and figure 3 of Ref. 28 are all good examples. Because the field distribution in Fig. 22 is for a surface mode, decaying fast, and the discontinuity between the uniform and periodic dielectric regions is not aligned with the zero of the field, the decaying field period length does not seem so apparent as the rest. (Note: only the distribution in the periodic dielectric region is decaying periodic solution.)

(2). *Numerical example for decaying full-period Floquet solutions* ($\Lambda$). Figure 23 of Ref. 15 is a good example for the 1D planar photonic crystal fibers. This figure is for the field distribution of a higher-order surface mode, decaying fast but showing more clear zeros. The period length of the decaying periodic solution can be obtained by measuring the distance between the first and third zeros of the field in the periodic dielectric region.

(3). *Numerical example from honeycomb fibers for decaying period solutions.* Another kind of photonic crystal fibers is made up of a hollow core surrounded by an array of microscopic air holes, or honeycomb optical fibers that is two-dimensional [9]. In such a structure, the field distribution in the band gap also looks like a 2D decaying periodic solution, as shown in Fig. 10 of the paper by Lin [29]. In this figure, the decaying field distribution shows two periods both along the $x$-axis (from 1.2 to 4.6) and along the $y$-axis (from 1.5 to 5.5) while the honeycomb structure has two periods along the $x$-axis and four periods along the $y$-axis. That means that the field distribution on the $x$-axis corresponds to a decaying full-period solution ($\Lambda$) while the distribution on the $y$-axis corresponds to a decaying half-period solution ($2\Lambda$).

To better understand the property of field distributions mentioned above, numerical examples for $TE_{ml}$ modes, with $m$ the mode index and $l$ the band index, are given in Figs. 1, 2, 3, and 4. Note: the number of zeros of $\partial H_x / \partial x$ in the guiding region is used to define the mode index for TE wave. For a $TE_{ml}$ mode, $H_x$ has $m$ extremums (peaks), and has an opposite parity to $m$; namely, $H_x$ is an odd (even) function of $x$ when $m$ is even (odd). As the guiding region width increases, fast-wave or guided mode dispersion curves, which are above the light line, shift down towards low frequency side to keep the same numbers of peaks in the guiding region. Due to the property of hyperbolic functions, high-index modes with $m \geq 2$ never intersect the light line.

It is seen from Figs. 2 and 4 that $H_x$ has double the period of the medium ($2\Lambda$) in the periodic structure both for $TE_{31}$ mode with $l = 1$ and $TE_{12,3}$ mode with $l = 3$, because their band indices are odd. In contrast, as shown in Fig. 3, $H_x$ has the same period as the medium ($\Lambda$) for the even band-index mode $TE_{82}$ with $l = 2$.

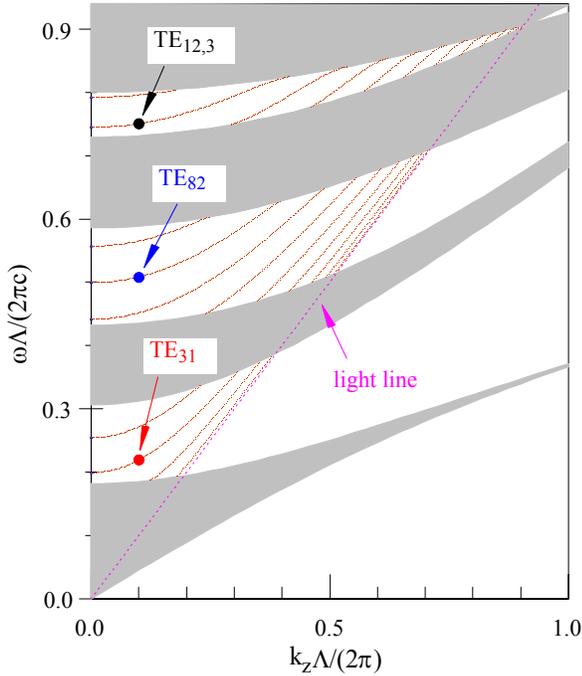

Fig. 1. TE-mode structure for a 1D PBG fiber consisting of a guiding region and two symmetric half-infinite bilayer periodical structures. White: stop band; gray: allowed band. Layer's refractive indices: $n_1 = 4.6$, $n_2 = 1.6$; layer's thickness: $\Lambda_1 = 0.3$ μm, $\Lambda_2 = 2.15$ μm; refractive index period: $\Lambda = 2.45$ μm. Vacuum guiding region width: $\Lambda_0 = 19.6$ μm ($8\Lambda$). The refractive index of the layers adjacent to the guiding region is $n_1$. Modes $TE_{31}$ (red), $TE_{82}$ (blue), and $TE_{12,3}$ (Black) are, respectively, located in the first, second, and third stop bands, and their field distributions at the marked frequency locations are shown in Figs. 2, 3, and 4 respectively.

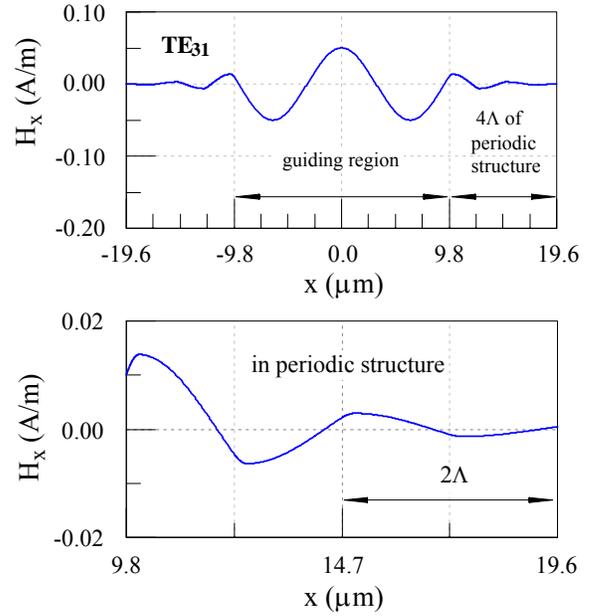

Fig. 2. Magnetic field $H_x$-distribution for $TE_{31}$ mode (odd band index). Top: $H_x$ is an even function and has three extremums (peaks) in the guiding region; bottom: $H_x$-period in the periodic structure is $2\Lambda$ (double the medium period). (All fields are drawn for a flowing power of 10 μW/m.)

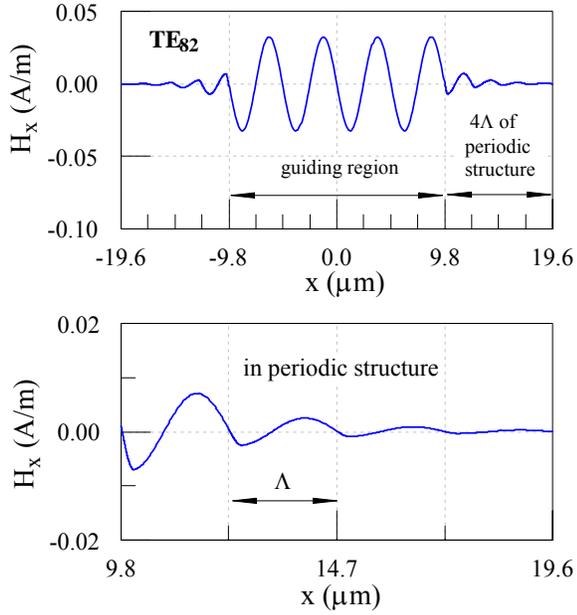

Fig. 3. Magnetic field $H_x$-distribution for $TE_{82}$ mode (even band index). Top: $H_x$ is an odd function and has eight extremums in the guiding region; bottom: $H_x$-period in the periodic structure is $\Lambda$ (the same as the medium period).

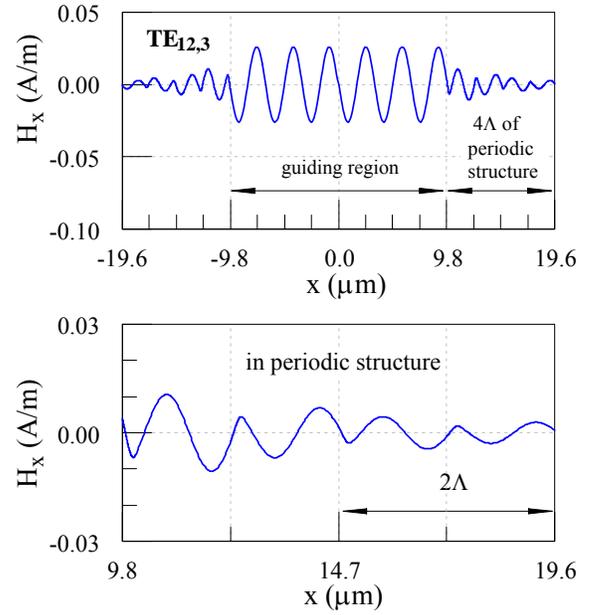

Fig. 4. Magnetic field $H_x$-distribution for $TE_{12,3}$ mode (odd band index). Top: $H_x$ is an odd function and has twelve extremums in the guiding region; bottom: $H_x$-period in the periodic structure is $2\Lambda$ (double the medium period).

Finally, it is interesting to point out that, like in the index guidance of conventional optical fibers, total internal reflection may occur when light of incidence irradiates on a half-infinite periodic structure from an empty guiding region. This is shown below. As indicated above, real $\sigma \neq \pm 1$ holds in a band gap, and real $k_x^{(n)} \neq 0$ is valid in any layer of the half-infinite periodic structure for an incident light from the empty region to form a guided mode. Consequently, we have the reflectivity $b_n b_n^* / a_n a_n^* = 1$ from Eqs. (17) or (18), which holds for every layer in the periodic structure, namely total internal reflection. In this sense, the ideal PBG guidance is also a form of total internal reflection. However, there are important differences between the two total internal reflections. In the index guidance, for example, the total internal reflection takes place when the incident angle is *larger* than a critical angle, while in the PBG guidance the total internal reflection for a common frequency range takes place when the incident angle is *less* than a critical angle, although this angle can be increased to 90 degrees so that omnidirectional-incidence total internal reflection is reached provided that the dielectric layers' refractive indices are large enough [30,31].

## 5. SUMMARY

The 1D PBG fiber actually is a transversely periodical boundary-value problem in classical electromagnetic theory, where two perfect conducting planes of the parallel-plate waveguide are replaced by two symmetric periodic structures. Ray-optics method based on transverse resonance condition, although it cannot be used to obtain slow-wave modes to provide an overview of the whole mode structure, is an alternative method for finding fast-wave modes, or guided modes; however, there is some inconvenience for this method, because the dispersion property of PBG mirrors makes the transverse resonance condition and mode index not one-to-one corresponding [32]. Field-solution method by matching boundary conditions is a strict method for solving 1D PBG fibers. Based on a real translation-matrix method, we have derived a general reflection coefficient expression, which is exactly a description of the PBG-mirror boundary conditions, and this expression can be directly applied for obtaining general dispersion equations for all electromagnetic modes, such as a mode that may provide single-mode guidance [32].

Simple and fundamental properties of Floquet solutions (Bloch waves), which characterize light field distributions in the band gap, are analytically shown, and they are identified in the numerical illustrations by a number of previously published research works. The properties revealed here are general and are applicable to any models of 1D planar periodic structures, no matter how many layers a unit cell has, what the contrast of refractive indices is, and the dielectric parameters in individual layers are homogeneous or inhomogeneous. Furthermore, we have shown that similar to the index guidance of conventional optical fibers, the PBG guidance is also a form of total internal reflection.

**REFERENCES**

1. J. D. Joannopoulos, P. R. Villeneuve, and S. Fan, "Photonic crystals: putting a new twist on light," Nature **386**, 143-149 (1997).
2. P. Russell, "Photonic crystal fibers," Science **299**, 358-362 (2003).
3. J. C. Knight, "Photonic crystal fibres," Nature **424**, 847-851 (2003).
4. A. F. Abouraddy, M. Bayindir, G. Benoit, S. D. Hart, K. Kuriki, N. Orf, O. Shapira, F. Sorin, B. Temelkuranl, and Y. Fink, "Towards multimaterial multifunctional fibres that see, hear, sense and communicate," Nature Photonics **6**, 336-347 (2007).
5. E. Yablonovitch, "Inhibited spontaneous emission in solid-state physics and electronics," Phys. Rev. Lett. **58**, 2059-2062 (1987).
6. S. John, "Strong localization of photons in certain disordered dielectric superlattices," Phys. Rev. Lett. **58**, 2486-2489 (1987).
7. P. Yeh, A. Yariv, and E. Marom, "Theory of Bragg fiber," J. Opt. Soc. Am. **68**, 1196-1201 (1978).
8. M. Ibanescu, Y. Fink, S. Fan, E. L. Thomas, and J. D. Joannopoulos, "An all-dielectric coaxial waveguide," Science **289**, 415-419 (2000).
9. T. A. Birks, P. J. Roberts, P. St. J. Russell, D. M. Atkin, and T. J. Shepherd, "Full 2-D photonic bandgaps in silica/air structures," Elect. Lett. **31**, 1941-1943 (1995).
10. R. F. Cregan, B. J. Mangan, J. C. Knight, T. A. Birks, P. St. J. Russell, P. J. Roberts, and D. C. Allan, "Single-mode photonic band gap guidance of light in air," Science **285**, 1537-1539 (1999).
11. B. Temelkuran, S. D. Hart, G. Benoit, J. D. Joannopoulos, and Y. Fink, "Wavelength-scalable hollow optical fibres with large photonic bandgaps for $CO_2$ laser transmission," Nature **420**, 650-653 (2002).
12. S-Y Lin, E. Chow, V. Hietala, P. R. Villeneuve, and J. D. Joannopoulos, "Experimental demonstration of guiding and bending of electromagnetic waves in a photonic crystal," Science **282**, 274-276 (1998).
13. S. A. Rinne, F. G-Santamaria, and P. V. Braun, "Embedded cavities and waveguides in three-dimensional silicon photonic crystals," Nature Photonics **2**, 52-56 (2008).
14. P. Yeh and A. Yariv, "Bragg reflection waveguides," Opt. Commun. **19**, 427-430 (1976).
15. P. Yeh, A. Yariv, and Chi-Shain Hong, "Electromagnetic propagation in periodic stratified media. I. General theory," J. Opt. Soc. Am. **67**, 423-438 (1977).
16. Zhi-fang Sang and Zhen-ya Li, "Optical properties of one-dimensional photonic crystals containing graded materials," Opt. Commun. **259**, 174-178 (2006).
17. P. C. Pandey, K. B. Thapa, and S. P. Ojha, "Large forbidden band in one-dimensional multi-layered structure containing exponentially graded materials," Opt. Commun. **281**, 1607-1614 (2008).
18. Changbiao Wang, "Simulation analysis of rectangular dielectric-loaded traveling wave amplifier for THz sources," Phys. Rev. ST Accel. Beams **12**, 120701(1)-(15) (2007).
19. Changbiao Wang and J. L. Hirshfield, "Theory for wakefields in a multizone dielectric lined waveguide," Phys. Rev. ST Accel. Beams **9**, 031301(1)-(18) (2006).
20. Wang Zhuxi and Guo Dunren, *Special functions* (Academic Press, Beijing, 1979).
21. E. T. Whittaker and G. N. Watson, *A course of modern analysis* (Cambridge, 1952).
22. J. Lekner, "Light in periodically stratified media," J. Opt. Soc. Am. A **11**, 2892-2899 (1994).
23. M. Born and E. Wolf, *Principles of optics* (*5th edition*) (Oxford, 1975).
24. K. F. Riley, M. P. Hobson, and S. J. Bence, *Mathematical methods for physics and engineering* (Cambridge, 2006).
25. A. Y. Cho, A. Yariv, and P. Yeh, "Observation of confined propagation in Bragg waveguides," Appl. Phys. Lett. **30**, 471-472 (1977).
26. G. Blanch, "Mathieu Functions," in *Handbook of mathematical functions with formulas, graphs, and mathematical tables*, M. Abramowitz and I. A. Stegun, eds. (Dover Publications, Inc., 1972), pp. 721-750.
27. A. Mizrahi and L. Schächter, "Optical Bragg accelerators," Phys. Rev. E **70**, 016505(1)-(21) (2004).
28. Zhiyu Zhang, S. G. Tantawi, and R. D. Ruth, "Distributed grating-assisted coupler for optical all-dielectric electron accelerator," Phys. Rev. ST Accel. and Beams **8**, 071302(1)-(8) (2005).
29. X. E. Lin, "Photonic band gap fiber accelerator," Phys. Rev. ST Accel. and Beams **4**, 051301(1)-(7) (2001).
30. J. N. Winn, Y. Fink, Shanhui Fan, and J. D. Joannopoulos, "Omnidirectional reflection from a one-dimensional photonic crystal," Opt. Lett. **23**, 1573-1575 (1998).
31. Changbiao Wang, "Simulation analysis of a highly omnidirectionalfar-infrared octaveband dielectric reflector," ShangGang Report #02-2008.
32. Changbiao Wang, "Single-mode propagation of light in one-dimensional all-dielectric light-guiding systems," Submitted for publication.

**Appendix**: Ref. 31 of the above paper, ShangGang Report #02-2008, is provided below.

# Simulation analysis of a highly omnidirectional far-infrared octaveband dielectric reflector

Changbiao Wang
*ShangGang Group, 70 Huntington Road, Apt. 11, New Haven, CT 06512, USA*
*changbiao_wang@yahoo.com*

A simulation analysis is given for a highly omnidirectional far-infrared octaveband dielectric reflector, which has nearly the same photonic band gap at different angles of incidence of light. It is shown that the omnidirectionality of a reflector fundamentally depends on the shape of the common area of stop band and incident-light band, and is also affected by the number of unit cells that the reflector contains. Band gap ratio and bandwidth ratio are introduced to describe the omnidirectionality.

Keywords: reflector, far-infrared, photonic band gap, ferroelectric

Omnidirectional dielectric reflectors based on one-dimensional photonic band gap (PBG) effect have been analyzed in the pioneering works [1,2], where each unit cell (period) consists of two layers of uniform dielectrics. To form a sharp contrast and increase reflectivity [3], one of the two layers has a higher refractive index while the other is set to have an index as low as possible, provided that the ambient-medium light line is above the Brewstwer's line to constitute a reasonable common band gap for all angles of incidence of light [1]. To improve the band gap, various approaches are employed [4,5]. However, if the index of the lower-index layer is not large enough, although a higher contrast can be reached, the band gap strongly depends on the angle of incidence [2], and the common band gap between normal and oblique incidences will be considerably reduced as the angle increases, resulting in deterioration of omnidirectionality.

There are two basic requirements on an ideal omnidirectionality: (*i*). The reflectivity (light power density reflection coefficient) is equal to unity at any angle of incidence for a certain frequency range, and (*ii*). The reflectivity frequency range does not depend on the incident angle — this is significant for applications to control of radiation modes. To improve the omnidirectionality, high-index materials are needed. Ferroelectric materials, as demonstrated by experiments, have attractive dielectric properties in the far-infrared waveband (0.1 THz ~ 1 THz) [6,7,8], and their refractive index can be made as low as 3 and as high as 19, which promisingly provides a possibility to realize highly omnidirectional reflectors. Based on such developing high-index materials, in this paper a far-infrared omnidirectional dielectric reflector is analyzed in simulations, which has nearly the same band gap at different angles of incidence and has more than one octave bandwidth.

For one-dimensional periodical dielectric structure with a low refractive-index contrast as an omnidirectional reflector, it has been shown that the index of layer's dielectric should be larger than that of the ambient medium [1]. That is true even for a high index-contrast structure, as shown in Fig. 1, where the index contrast is 18, with a quarterwave stack taken. It is seen from Fig. 1 that there is a green zone that is a common area of the incident-light band and the structure stop band. It can be shown that the incident light with a wave vector within the green zone gets an exactly total reflection from a half-infinite periodic reflector. Although such a total reflection never can be obtained from a finite periodic reflector, the reflectivity can be very close to unity. Therefore, the omnidirectionality of a reflector fundamentally depends on the shape of green zone, and is also affected by the number of unit cells that the reflector contains. In Fig. 1, because the index of the ambient medium is the same as one of the layer's indices, part of the green-zone boundary is almost tangential to the light line and more wave vectors of low-frequency light fall out of the green zone as the incident angle increases, resulting in the reflector working only in a limited range of angle [2].

Figure 2 shows the band structure for a bilayer periodical structure with layer's refractive indices $n_1 = 18$ and $n_2 = 5$ and layer's thickness $\Lambda_1 = 8$ μm and $\Lambda_2 = 28.8$ μm. It is seen that the light line of TM wave is far above the Brewster's line and the green zone looks like an upside-down isosceles trapezoid, although the index contrast is reduced to 18/5 = 3.6 compared with 18/1 = 18 in Fig. 1. From the maximum and minimum values of the normalized frequency $\omega\Lambda/(2\pi c)$ on the top and bottom boundaries of the green zone, we can obtain the *band gap ratio*, given by $(B_{t\min} - B_{b\max})/(B_{t\max} - B_{b\min}) = 0.96$; obviously, the band gaps are nearly the same for different incident angles.

Figure 3 shows the reflectivity of a far-infrared octaveband reflector that contains five unit cells with layer's parameters in Fig. 2. Suppose that a wavelength range with a reflectivity of larger than 0.99, corresponding to > 20-dB transmittance attenuation, is taken to be bandwidth. It is found that the common bandwidth of this reflector is more than one octave bandwidth, starting from 418 μm to 895 μm. The maximum bandwidth is 558 μm at normal incidence, while the minimum bandwidth is 477 μm at 80-degree incidence, with a *bandwidth ratio* of 477/558 = 0.85, considerably less than the band gap ratio 0.96. This is because this reflector only has five periods. The band gap ratio and the bandwidth ratio are different measures: the band gap ratio is decided by the stop band and incident-light band, while the bandwidth ratio additionally depends on the number of periods (unit cells) that a specific reflector has.  In other words, a good band gap ratio is a necessary condition to obtain highly omnidirectional reflectors, while the bandwidth ratio is a direct measure of omnidirectionality.  In principle, the two ratios get closer as the number of unit cells increases for a band structure with a good shape of green zone.  Calculations show that the bandwidth ratio is increased to 0.95 when the number of unit cells is increased to ten from five, well close to the band gap ratio 0.96.

It should be pointed out that material-caused losses in the above numerical calculations are ignored to simplify the analysis.  Like in all electronic devices, such losses will certainly affect the reflector's performance [9].  Existing ferroelectrics have considerable losses in the THz region [6]; however, as the development of material technology, material properties could be further improved.

In summary, a simulation analysis of a far-infrared highly omnidirectional octaveband reflector has been presented.  Because of its good omnidirectionality, this reflector could be used to control the selection of harmonics in far-infrared radiation generators, such as coherent synchrotron radiators [10].  In addition, the dielectric constant of ferroelectric materials is electrically adjustable, and this reflector might have some potential applications to electrically-controlled switches and filters in THz-technology field.

Optics in periodically stratified medium is an old subject, and it can be traced back to about one hundred years ago [11].  However, it was not until the late 1970's that one found that such periodic structures could be used to guide light in an empty channel [12].  Unfortunately, Yeh, Yariv, and Hong did not realize then that the periodic structures also could be used for omnidirectional reflectors although the illustrations of their TE- and TM-band structures implicitly include this property [12].  More than twenty years later, it was Winn, Fink, Fan, and Joannopoulos who exposed this property [1], with experimental demonstrations subsequently [2].  Recently, it has been shown theoretically that there are two kinds of light field distributions in a band gap [13]; interestingly, this distribution property was also identified in the numerical examples of Ref. 12.  As the advancement of material science and creation of metamaterials, investigations on this old subject have become more intensive and extensive, with more new features kept being dug up.

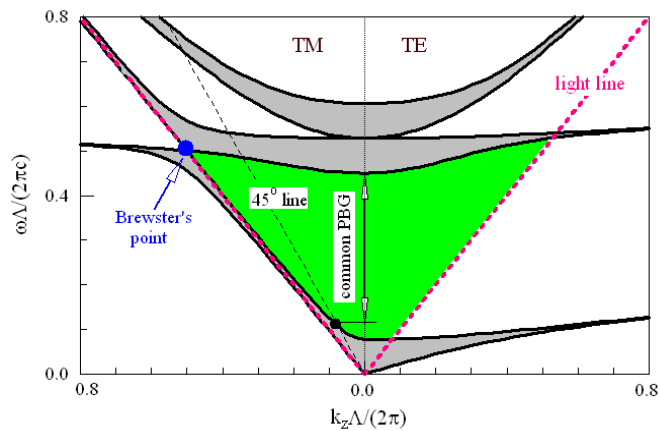

Fig. 1.  TM- and TE-wave band structure in $\omega\Lambda/(2\pi c) - k_z\Lambda/(2\pi)$ plane for a bilayer periodical structure, with light line *very close to* but *below* Brewster's line, where $\omega$ is the light frequency, $k_z$ is the axial wave number, $c$ is the vacuum light speed, and $\Lambda$ is the structure period.  Layer's refractive index: $n_1 = 18$ and $n_2 = 1$; layer's thickness: $\Lambda_1 = 8$ μm and $\Lambda_2 = 144$ μm.  Dark zones are allowed bands and white zones are stop bands; and the light cone formed by two red dashed light lines is the incident-light band – a collection of $(\omega/c, k_z)$ of incident light from the ambient medium (air).  The green zone is the first common area of the stop band and the incident-light band.  "Common PBG" in the figure denotes the common band gap between normal and 45-degree incident angles, and such a common band gap is rapidly reduced as the incident angle increases because the light line is almost tangential to the stop-band boundary there.

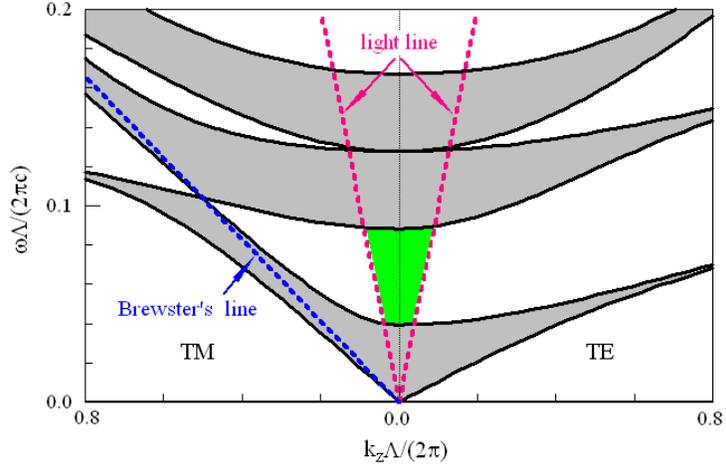

(a). Band structure.

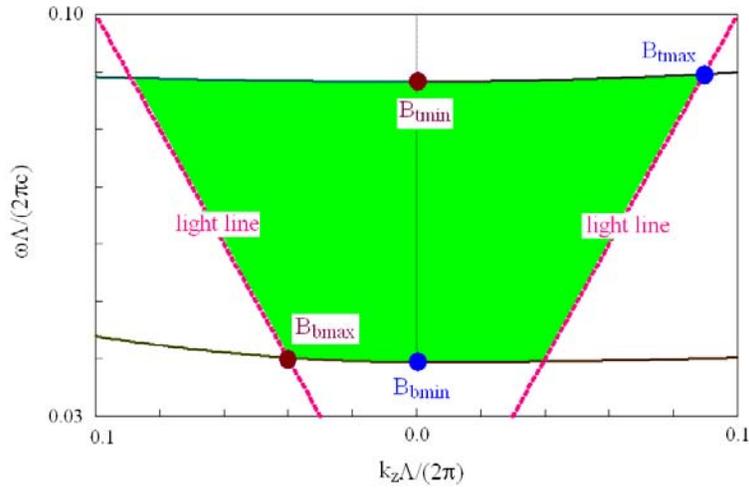

(b). Magnified green zone.

Fig. 2. (a). TM- and TE-wave band structure for a bilayer periodical structure with light line *well above* Brewster's line. Layer's refractive index: $n_1 = 18$ and $n_2 = 5$; layer's thickness: $\Lambda_1 = 8$ μm and $\Lambda_2 = 28.8$ μm. Note: the green zone is flush bordered by left and right light lines, and bounded by nearly parallel top and bottom stop-band boundaries. (b). Magnified green zone. The band gap ratio is 0.96.

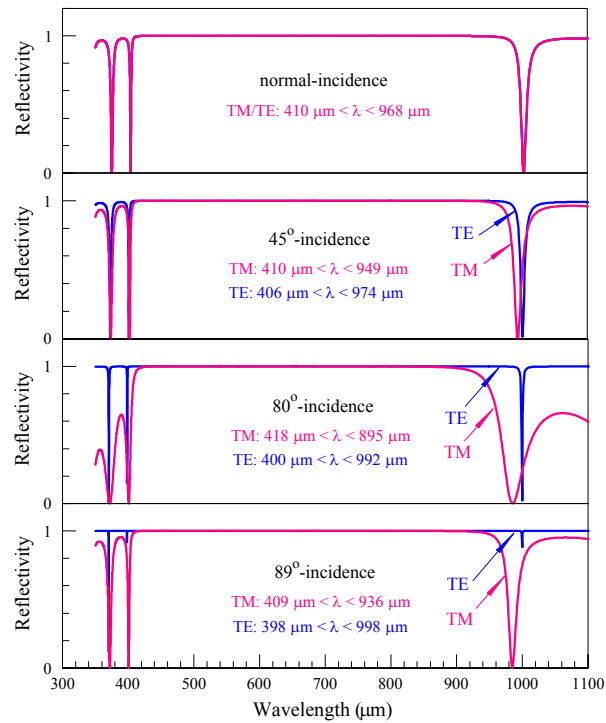

Fig. 3. TM- and TE-wave reflectivity for a far-infrared reflector with five unit cells with layer's parameters in Fig. 2. The largest bandwidth is 558 μm at the normal incidence, and the smallest bandwidth is 477 μm at the 80-degree incidence, with a bandwidth ratio of 0.85. Note: the reflectivity profiles for TM and TE waves at normal incidence are the same, and λ denotes the wavelength in free-space.

## References

1. J. N. Winn, Y. Fink, Shanhui Fan, and J. D. Joannopoulos, Opt. Lett. **23**, (1998) 1573.
2. Y. Fink, J. N. Winn, Shanhui Fan, Chiping Chen, J. Michel, J. D. Joannopoulos, and E. L. Thomas, Science **282**, (1998) 1679. Note: the 9-layer reflector in this reference uses a stack of dielectric films with refractive indices 4.6 and 1.6 and a thickness ratio of 0.8 μm to 1.65 μm. For the convenience of comparison, I repeated this example to obtain some data that are not directly given in this reference. The maximum reflectivity of TM wave at 80°-incidence is less than 0.99. For a 0.98-reflectivity definition (17-dB transmittance attenuation), the TE-TM common gaps for normal, 80°-, and 89°-incidence are, respectively, 8.9 μm, 3.1 μm, and 2.9 μm; lying from 9.8 to 18.7 μm, 9.8 to 12.9 μm, and 8.7 to 11.6 μm. Thus the omnidirectional-incidence common gap is about (11.6-9.8 =) 1.8 μm, with a bandwidth ratio of (11.6-8.7)/(18.7-9.8) = 0.33.
3. M. Born and E. Wolf, "Principles of optics (5$^{th}$ edition)," (Oxford, London, 1975).
4. Xin Wang, Xinhua Hu, Yizhou Li, Wulin Jia, Chun Xu, Xiaohan Liu, and Jian Zi, Appl. Phys. Lett. **80**, (2002) 4291.
5. Peng Han and Hezhou Wang, J. Opt. Soc. Am. B **20**, (2003) 1996.
6. Jiaguang Han, Fan Wan, Zhiyuan Zhu, and Weili Zhang, Appl. Phys. Lett. **90**, (2007) 031104. This reference reports that the different refractive indices (3 and 18) of the powder form of ferroelectric $SrTiO_3$ are obtained by controlling its purity.
7. M. Misra, K. Kotani, I. Kawayama, H. Murakami, and M. Tonouchi, Appl. Phys. Lett. **87**, (2005) 182909.
8. P. Kužel, F. Kadlec, H. Němec, R. Ott, E. Hollmann, and N. Klein, Appl. Phys. Lett. **88**, (2006) 102901.
9. M. Skorobogatiy and A. Dupuis, Appl. Phys. Lett. **90**, (2007) 113514.
10. Changbiao Wang, Phys. Rev. A **38**, (1988) 6215.
11. J. W. S. Rayleigh, Proc. R. Soc. London Ser. A **93**, (1917) 565.
12. P. Yeh, A. Yariv, and Chi-Shain Hong, J. Opt. Soc. Am. **67**, (1977) 423. Note: in this reference, Figs. 3 and 4 show TE- and TM-band structures respectively for a periodic structure of a stack of dielectric films with refractive indices 3.38 and 2.89, and the two figures also implicitly show a small common gap [$\Delta(\omega\Lambda/c) = 0.013\pi$] in the first common area of TE-TM stop band and incident-light band (not drawn). This can be seen as follows. The gaps for TE and TM waves at normal incidence overlap, and the common gap of TM wave is smaller than the one of TE wave. Therefore, the TM common gap ($0.013\pi$) is the common gap of TE and TM waves.
13. Changbiao Wang, "Light Field Distributions in One-dimensional Photonic Crystal Fibers," submitted for publication in Journal of the Optical Society of America B **26**, 603-609 (2009).